\documentclass{aa}

\usepackage{graphicx}
\usepackage{txfonts}
\usepackage{natbib,twoopt}
\usepackage{longtable}
\usepackage[breaklinks=true]{hyperref} 
\bibpunct{(}{)}{;}{a}{}{,} 
\newcommandtwoopt{\citeads}[3][][]{\href{http://adsabs.harvard.edu/abs/#3}%
{\citealp[#1][#2]{#3}}}
\newcommandtwoopt{\citepads}[3][][]{\href{http://adsabs.harvard.edu/abs/#3}%
{\citep[#1][#2]{#3}}}
\newcommandtwoopt{\citetads}[3][][]{\href{http://adsabs.harvard.edu/abs/#3}%
{\citet[#1][#2]{#3}}}
\newcommandtwoopt{\citeyearads}[3][][]%
{\href{http://adsabs.harvard.edu/abs/#3}{\citeyear[#1][#2]{#3}}}
\begin{document}

   \title{Searching for chemical inhomogeneities in Open Clusters}

   \subtitle{Analysis of the CN and CH Molecular Band Strengths in NGC~2158, NGC~2420, NGC~2682, NGC~7789 and Berkeley~29}

   \author{R. Carrera\inst{1,2}
                \and
               C. E. Mart\'{\i}nez-V{\'a}zquez\inst{1,2}}

   \institute{Instituto de Astrof\'{\i}sica de Canarias, La Laguna, Tenerife, Spain\\
             \email{rcarrera@iac.es; cmartinez@iac.es}
             \and
             Departamento de Astrof\'{\i}sica, Universidad de La Laguna, Tenerife, Spain\\
             }

   \date{Received September 15, 1996; accepted March 16, 1997}

\titlerunning{Searching for multiple stellar populations in Open Clusters}
\authorrunning{Carrera \& Mart\'{\i}nez-V{\'a}zquez}

  \abstract
   {The total mass of a cluster, being the main parameter determining its ability to host more than one
   stellar generation, may constitute a  threshold below which the cluster is able to form only a single stellar population.}
   {Our goal is to investigate the existence of star-to-star variations of CN and CH band strengths,
   related to the N and C abundances, respectively, among the stars in five open cluster (NGC~2158, NGC~2420,
   NGC~2682, NGC~7789 and Berkeley~29) similar to those observed in globular clusters and linked with the
   existence of multiple populations therein. Since these systems are less massive than globulars, our
   results may allow us to constrain the lower mass necessary to form more than one stellar population.}
   {We measured the strength of the CN and CH bands, which correlate with the C and N abundances,
   using four molecular indices in low-resolution SDSS/SEGUE spectra.}
   {We found that for four of the open clusters (NGC~2158, NGC~2420, NGC~2682 and Berkeley~29)
   all the stars studied in each of them have similar CN and CH band strengths within the uncertainties
   since neither anomalous spreads nor bimodalities have been detected in their CN and CH distributions.
   In contrast, for NGC~7789 we found an anomalous spread in the strength of the CN molecular band at
   3839 \AA~which is larger than the uncertainties. However, the small number of stars studied in this
   cluster implies that further analysis is needed to confirm the existence of chemical inhomogeneities in this cluster.}
      {}


   \maketitle
%

\section{Introduction}\label{sec1}

It seems that almost all Galactic globular clusters (GCs), properly studied, host multiple stellar populations
that explain the previously known star-to-star variations of the chemical abundances of light elements such
as C, N, Na, O, Al and Mg \citepads[e.g.][and references therein]{2012A&ARv..20...50G}. Similar features have
been reported in extragalactic GCs \citepads[e.g.][]{2006A&A...453..547L,2009ApJ...695L.134M}. Several main-sequence (MS) turn-offs or anomalous spreads in this region have been observed in intermediate-age and young massive clusters in the Magellanic Clouds \citepads[e.g.][]{2008AJ....136.1703G,2009A&A...497..755M}. Although several hypotheses have been proposed to explain these findings, the most accepted one is the so-called self-enrichment scenario first presented by  \citetads{1980ApJ...236L..83D}. According to this hypothesis, a stellar cluster has experienced two or more star formation episodes, or extended periods of star formation. Therefore, the subsequent stellar populations born from the interstellar medium that have been polluted by material ejected from previous generations. This material may have been synthesized through the CNO cycle so that the younger populations have become enhanced in He, N, Na and Al and depleted in C, O, Ne and Mg. Although the main progenitors of this polluted material are still unclear, the most accepted candidates are either intermediate-age asymptotic giant branch (AGB) stars \citepads[e.g.][]{2007MNRAS.379.1431D} or fast rotating massive stars \citepads[e.g.][]{2007A&A...464.1029D}.

\begin{figure}
\centering
\includegraphics[width=\hsize]{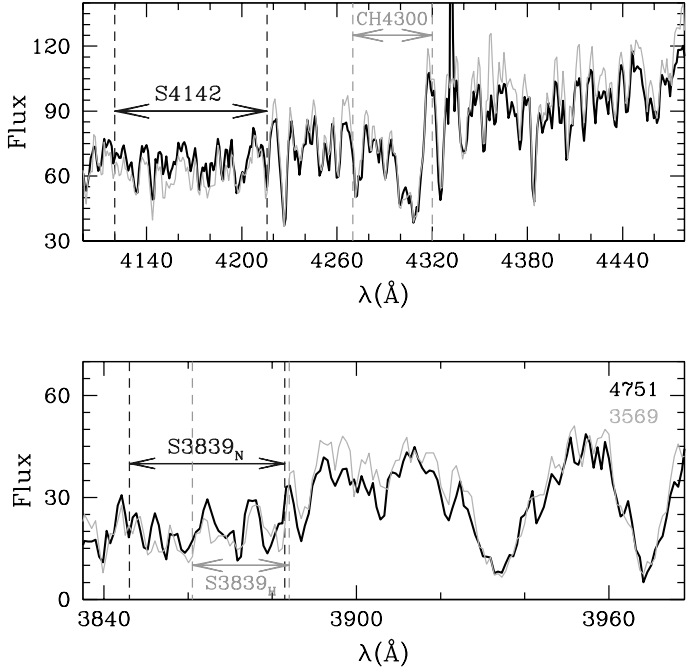}
\caption{Example of two spectra over-plotted, one in grey (3569) and one in black (4751), for two stars in NGC~7789 with different CN band strengths. The top panel shows the wavelength region which includes the CN band at 4142 \AA~and the CH band at 4300 \AA. The wavelength region which contains the CN band at 3839 \AA~has been plotted in the bottom panel. The windows used to determine the strength of each molecular band have been marked with horizontal arrows whereas vertical dashed lines shows the edge of these windows.\label{fig_spectra}}
\end{figure}

The ability of a stellar cluster to retain the polluted material from which the subsequent stellar populations
are formed may depend mainly on the gravitational potential of the cluster. Other variables such as age,
metallicity or environment may also play a role in this context. Therefore, a minimum mass may be required
to retain the material ejected by the first generations. From an observational point of view, this limit may lie between 10$^4$ and 10$^5$ M$_{\odot}$ \citepads[e.g.][]{2009ApJ...695L.134M,2010A&A...516A..55C}. However,
these values refer to present-day mass, and may have been higher at the moment of the cluster's formation since
it is expected that clusters will have lost a significant fraction of their initial mass, either in form of gas or stars,  during their lifetime \citepads{2008ApJ...685..247B,2010MNRAS.409..305L}. Theoretical investigations point to an
initial mass limit between $\sim$ 10$^5$ and 10$^6$ M$_{\odot}$ \citepads[e.g.][]{2010ApJ...718L.112V,2011MNRAS.412.2241B}.

Only a handful of GCs are known with masses lower than 10$^5$ M$_{\odot}$. Two systems in this mass range, Palomar~12 and Terzan~7, seem not to have star-to-star variations of light element abundances, although in the best case only five stars have been studied \citepads{2004AJ....127.1545C,2004AJ....127..373T,2005A&A...437..905S,2007A&A...465..815S}.
Similar results have been found by \citetads{2008A&A...486..437K} from the analysis of CN and CH molecular bands,
related to N and C abundances, respectively, in low-resolution spectra. However, \citetads{2010A&A...524A..44P}
analysed similar data in 23 stars of Palomar~12, finding a clear bimodal distribution of the CN band strengths and
anti-correlation with the CH band strengths. Star-to-star variations of Na and O abundances have been detected recently by \citetads{2012ApJ...756L..40G} in a sample of 21 giant stars in the old \citepads[$\sim$8 Gyr][]{2012A&A...543A.106B} and metal-rich ([Fe/H]$\sim$+0.40) open cluster (OC) NGC~6791 with a present-day mass similar to that of Palomar~12 and Terzan~7 \citepads[$10^4$ M$_{\odot}$,][]{2011ApJ...733L...1P}. \citetads[][hereafter Paper I]{2012ApJ...758..110C} found that the star-to-star variations of the strengths of the CN and CH molecular bands are observed not only in red giant branch (RGB) objects but also among MS stars. Moreover, NGC~6791 shows an unexpectedly wide RGB and MS, which could be related to the existence of an age spread of about 1 Gyr \citepads{2011ApJ...727L...7T}. However, they could also be explained in terms of the existence of differential line-of-sight reddening \citepads{2011ApJ...733L...1P}.
In contrast, chemical variations of light element abundances have not been observed in a sample of 30 stars in the
OC Berkeley~39 \citepads{2012A&A...548A.122B}. This cluster has a similar mass and age to that of NGC~6791 although it is more metal-poor ([Fe/H]$\sim$-0.21). Other investigations on OCs have also investigated the existence of star-to-star variations of light element abundances from high-resolution spectroscopy without success,
although the number of stars analysed in each cluster is rather small \citepads[e.g.][]{2009A&A...500L..25D,2009A&A...502..267S,2010A&A...511A..56P,2011A&A...535A..30C}.
The strengths of CN and CH molecular bands in low-resolution spectra have also been studied in several OCs
with the same results although only about ten stars were studied in the best of the cases
\citepads[e.g.][]{1984AJ.....89..263S,1985AJ.....90.2526N,2009PASP..121..577M}. Extended MS turn-offs have also
been found in two clusters, SL~529 and SL~862, in the Large Magellanic Cloud (LMC) with masses similar to
that of  NGC~6791, although they are younger \citepads{2009A&A...497..755M,2013MNRAS.430.2358P}. These features have also been interpreted in terms of the existence of an extended period of star formation in both clusters. However, chemical analyses are needed to confirm, or discard, the existence of the star-to-star abundance variations in the two clusters.

\begin{figure}
\includegraphics[width=\hsize]{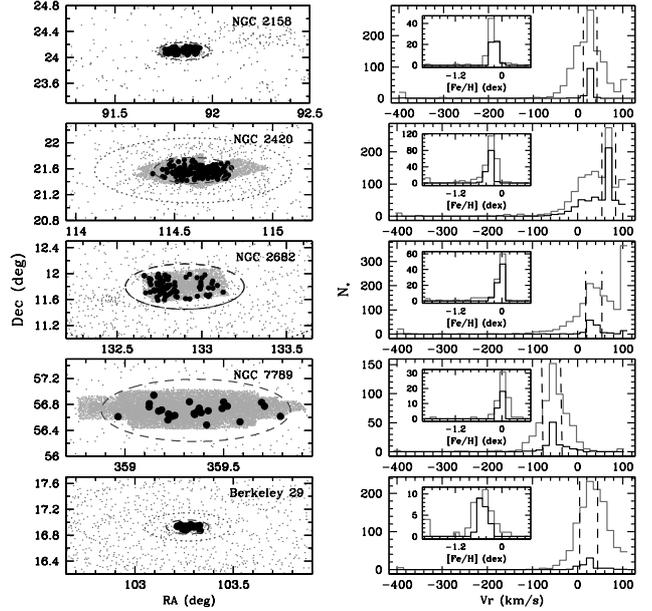}
\caption{Left: Spatial distribution of all stars observed by SEGUE (points) and those with available photometry
(light grey) for each cluster. Dashed lines denote the radii used to select cluster members, which are marked as
black dots. Right: Radial velocity histograms of all stars observed by SEGUE (grey) and those located within the
radii used to select cluster members (black). Dashed lines denoted the radial velocity limits used (see
Table~\ref{tbl-1}). [Fe/H] distributions of stars selected for their radial velocity and [Fe/H]
have been plotted in the inset panel with grey and black lines, respectively.\label{fig_sele}}
\end{figure}

\begin{table*}
\caption{Information about studied clusters.\label{tbl-1}}
\centering
\begin{tabular}{lccccccccc}
\hline\hline
Cluster & $\alpha_{2000}$ & $\delta_{2000}$ & Age & $\left\langle v_r\right\rangle$  & $[Fe/H]$ & Radius & Stars\tablefootmark{a} & Ref \\
  & hrs & deg & Gyr & km s$^{-1}$ & dex & arcmin & & \\
\hline
NGC 2158 & 06:07:25 & +24:05:48 & 1.0 & 24.9$\pm$1.9 & -0.15$\pm$0.14 & 8.0 & 35 & 1,3,4 \\
NGC 2420 & 07:38:23 & +21:34:24 & 2.0 & 69$\pm$5 & -0.39$\pm$0.20 & 12.0 & 67 & 1,3,5 \\
NGC 2682 & 08:51:18 & +11:48:00 & 2.6 & 36$\pm$6 & 0.01$\pm$0.07 & 21.0 & 46 & 1,3,6,7 \\
NGC 7789 & 23:57:24  & +56:42:30 & 1.4 & -58$\pm$6 & -0.04$\pm$0.05 & 28.84 & 11 & 1,3,6,7,8\\
Berkeley 29 & 06:53:18 & +16:55:00 & 1.1 & 24.7$\pm$0.4 & -0.42$\pm$0.07 & 7.5 & 9 & 1,2,3\\
\hline
\end{tabular}
\tablefoot{\tablefoottext{a}{This is the total number of stars used in our analysis for each cluster as a function of their spatial distribution, radial velocities, metallicities and position in the colour-magnitude diagram and with photometric magnitudes and good signal-to-noise ratio spectrum available (see text for details).}}
\tablebib{(1)~\citetads{2002A&A...389..871D}; (2)~\citetads{2008A&A...488..943S}; (3)~\citetads{2011A&A...535A..30C}; (4)~\citetads{2011AJ....142...59J}; (5)~\citetads{2004AJ....128.2306C}; (6)~\citetads{2007AJ....134.1298C};
(7)~\citetads{2002AJ....124.2693F}; (8)~\citetads{2007AJ....133.2061W}}
\end{table*}

It is essential to better constrain the minimum cluster mass required to form multiple stellar populations. For this purpose it is necessary to sample properly less massive systems by analyzing a statistically significant number of stars. Following the same procedure as used in Paper I, the goal of this paper is to analyse the strengths of the CN and CH molecular bands in stars at different evolutionary stages in five OCs: NGC~2158, NGC~2420, NGC~2682 (M~67), NGC~7789 and Berkeley~29. The last two have present-day masses similar to that of NGC~6791. Although
this technique does not supply chemical abundances directly, it does provide valuable information about the existence of star-to-star variations of C and N abundances. The observational material utilized in this paper is presented in Section~\ref{sec2}. The molecular indices used to determine the CN and CH band strengths are presented in  Section~\ref{sec3} together with the procedure used to obtain their distributions. The results obtained for each cluster are discussed in Section~\ref{sec4}. Finally, the main results are summarized in Section~\ref{sec5}.

\section{Observational Material}\label{sec2}

In the framework of the Sloan Extension for Galactic and Understanding and Exploration (SEGUE) survey
\citepads{2009AJ....137.4377Y}, low-resolution (R$\sim$2000) spectra were obtained for $\sim$240\,000
stars over a wavelength range of 3800--9200 \AA. These includes several globular and open clusters,
including those studied here, which were used to validate the SEGUE Stellar Parameter Pipeline
\citepads[SSPP, ][]{2008AJ....136.2070A,2008AJ....136.2022L,2008AJ....136.2050L}. The reduced spectra,
together with atmosphere parameters (T$_{eff}$, log~g, [Fe/H]), radial velocities, etc., derived for observed stars were made public in the eighth Sloan Digital Sky Survey (SDSS) data release
\citepads{2011ApJS..193...29A}. Two examples of the spectra analysed in this paper are shown in Figure~\ref{fig_spectra}. The procedure for obtaining this information can be summarized as follows. Raw spectra are first reduced by the SDSS spectroscopic reduction pipeline, described in detail by
\citetads{2002AJ....123..485S}, which provides flux- and wavelength-calibrated spectra, together with initial determinations of the radial velocities and spectral types. More accurate radial velocities are
calculated in a subsequent step by SSPP, together with determinations of metallicity, effective temperature
and surface gravity.

The same method used in Paper I and fully described by \citetads{2011AJ....141...89S}
 was used to constrain the members of each cluster. In brief, we firstly rejected those stars located further than a given distance from the cluster centre (left panels of Fig.~\ref{fig_sele}). The cluster centres and radius used in each case are listed Table~\ref{tbl-1}. In general, this distance has been chosen as half of the tidal radius of each
cluster.  Exceptional cases were NGC~2158 and NGC~7789 for which we used the tidal radius since it seems
that in both cases there are still cluster stars beyond the assumed tidal radius. In the next step, we used the
values provided by SSPP to discard those stars with radial velocities and [Fe/H] that were not within 3$\sigma$ of
the mean radial velocity and [Fe/H] assumed for each cluster and listed in Table~\ref{tbl-1} (right panels of
Fig.~\ref{fig_sele}). The total number of stars considered as cluster members from these criteria are 63, 115,
91, 29 and 31 for NGC~2158, NGC~2420, NGC~2682, NGC~7789 and Berkeley~29, respectively.

\begin{figure}
\centering
\includegraphics[width=\hsize]{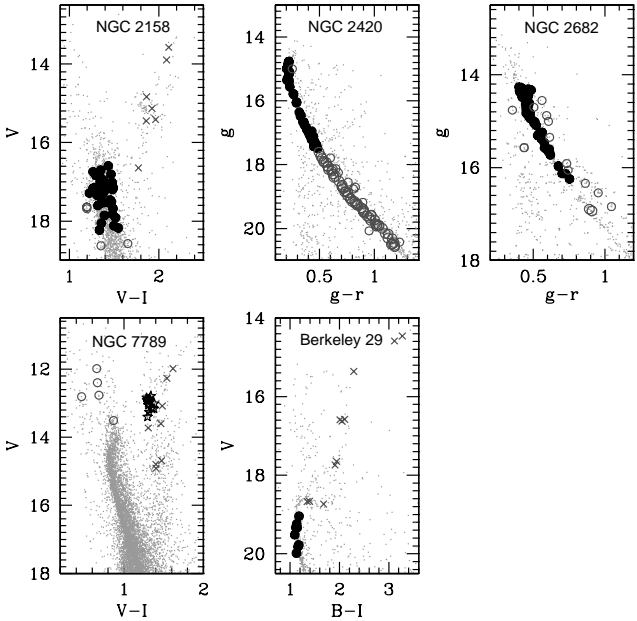}
\caption{Colour--magnitude diagrams of the five clusters in our sample with the location of the our target stars:
MS (filled circles); RGB (crosses), RC (open stars), and discarded stars either from their position in the
colour--magnitude diagram or by low signal-to-noise ratio spectra (open circles). The best photometries available in the literature have been used in spite this implies different filters combinations since similar quality photometry for all the clusters in our sample are not available in the same bandpasses.\label{fig_dcm}}
\end{figure}

Although SDSS data releases also provide information about the magnitudes of the target stars, we preferred to
use other sources since the SDSS photometric package is not able to properly handle high-density crowded fields
typical of the central areas of clusters, as denoted by \citetads{2008ApJS..179..326A}. For NGC~2420 and NGC~2682 we used the photometry obtained by \citetads{2008ApJS..179..326A} in Sloan $ugri$ bandpasses. For the other three systems (NGC~2158, NGC~7789 and Berkeley~29) we used the photometry obtained by \citetads{2002MNRAS.332..705C}, \citetads{1998PASP..110.1318G} and \citetads{2004MNRAS.354..225T}, respectively. Unfortunately, we have not been able to match all the stars selected as cluster members from the criteria described above with the photometric sources. The total number of stars that meet with these criteria, and that also have available photometric magnitudes are 47, 114, 74, 25 and 20  for NGC~2158, NGC~2420, NGC~2682, NGC~7789 and Berkeley~29, respectively. Finally, these stars are grouped as a function of their position in the colour-magnitude diagrams. Selected stars have been over-plotted to the colour-magnitude diagram of each clusters with different symbols as a function of their evolutionary stage in Figure~\ref{fig_dcm}: filled circles, crosses and open stars for MS, RGB and red clump (RC) stars, respectively. Stars discarded owing to their position in the colour--magnitude diagram or because their spectra have low signal-to-noise ratios have been plotted with open circles. Since not enough RGB stars have been observed in the clusters of our sample, we restricted our analysis to the MS stars with the exception of NGC~7789, in which the RC objects have been studied. The total number of stars used in each cluster are listed in column 8 of Table~\ref{tbl-1}.

\section{CN and CH band strenghs}\label{sec3}

\begin{table*}
\caption{Index measurements for the sample stars.\label{table_sample}}
\begin{tabular}{lccccccccccc}
\hline\hline
 Cluster & Plate\tablefootmark{a} & Fiber\tablefootmark{a} & R.A. & Dec. & V & g & ID$_{phot}$\tablefootmark{b} & $S3839_{H}$ & $S3839_{N}$ & $S4142$ &  $CH4300$\\
\hline
NGC~2158 & 2912 & 409 & 6:07:10.96 & +24:09:33.5 & 17.51 & & 899 & -0.50$\pm$0.03 & -0.28$\pm$0.03 & -0.32$\pm$0.03 & -0.48$\pm$0.03\\
NGC~2158 & 2912 & 416 & 6:07:14.18 & +24:08:08.8 & 17.53 & & 919 & -0.49$\pm$0.03 & -0.28$\pm$0.03 & -0.33$\pm$0.03 & -0.48$\pm$0.03\\
NGC~2158 & 2912 & 417  & 6:07:14.25 & +24:10:13.0 & 17.85 & & 1143 & -0.49$\pm$0.04 & -0.26$\pm$0.04 & -0.32$\pm$0.04 & -0.47$\pm$0.04\\
\hline
\end{tabular}
\tablefoot{\tablefoottext{a}{Plate and fiber uniquely identify the spectrum used in this analysis for each star in the SDSS III database.}
\tablefoottext{b}{ID$_{phot}$ identifies the star in the photometry used.}}

\end{table*}

\begin{table*}
\caption{Median uncertainties for each cluster and index and the $\sigma$ of the Gaussian fit to the
 generalized histogram of each pseudo-index.\label{table_median_uncertainties}}
\begin{tabular}{lcccccccc}
\hline\hline
 Cluster & \multicolumn{2}{c}{$S3839_{H}$} & \multicolumn{2}{c}{$S3839_{N}$} & \multicolumn{2}{c}{$S4142$} & \multicolumn{2}{c}{$CH4300$} \\
& $\langle \sigma \rangle$ & $\sigma_{Gaussian}$ & $\langle \sigma \rangle$ & $\sigma_{Gaussian}$ & $\langle \sigma \rangle$ & $\sigma_{Gaussian}$ & $\langle \sigma \rangle$ & $\sigma_{Gaussian}$ \\
\hline
NGC~2158 & 0.025$\pm$0.006 & 0.033$\pm$0.001 & 0.027$\pm$0.006 & 0.033$\pm$0.001 & 0.026$\pm$0.006 & 0.029$\pm$0.001 & 0.025$\pm$0.006 & 0.029$\pm$0.001\\
NGC~2420 & 0.030$\pm$0.008 & 0.034$\pm$0.002 & 0.031$\pm$0.009 & 0.035$\pm$0.002 & 0.030$\pm$0.008 & 0.031$\pm$0.001 & 0.030$\pm$0.008 & 0.031$\pm$0.001 \\
NGC~2682 & 0.034$\pm$0.003 & 0.041$\pm$0.000 & 0.036$\pm$0.003 & 0.042$\pm$0.000 & 0.033$\pm$0.002 & 0.034$\pm$0.000 & 0.033$\pm$0.005 & 0.034$\pm$0.000 \\
NGC~7789 & 0.056$\pm$0.002 & 0.079$\pm$0.004 & 0.061$\pm$0.002 & 0.089$\pm$0.004 & 0.050$\pm$0.001 & 0.058$\pm$0.003 & 0.048$\pm$0.001 & 0.049$\pm$0.002\\
Berkeley~29 & 0.09$\pm$0.02 & 0.12$\pm$0.02 & 0.09$\pm$0.02 & 0.12$\pm$0.02 & 0.09$\pm$0.02 & 0.10$\pm$0.01 & 0.09$\pm$0.02 & 0.10$\pm$0.01\\
\hline
\end{tabular}
\end{table*}

The strengths of the CN molecular bands at 3839 and 4142 \AA~and the CH band at $\sim$4300 \AA~were determined using the same indices as defined in Paper I. We refer the reader to this paper and references therein for a detailed discussion about each index. In brief, we used a two index definitions for the CN band at 3839 \AA~by \citetads{1981ApJ...244..205N} and \citetads{2003AJ....125..197H}. Each is optimized to sample RGB and MS stars respectively. For the other two molecular bands, the CN one centred at 4142 \AA~and the CH one at 4300 \AA~we used the index definitions by
\citetads{1979ApJ...230L.179N} and \citetads{1999AJ....118..920L}, respectively. As in Paper I, the uncertainty in each index was calculated assuming pure photon (Poisson) noise statistics in the flux measurements as in
\citetads{2010A&A...524A..44P}. The indices and errors determined together with photometric magnitudes for each
selected star are listed in Table~\ref{table_sample}. The median uncertainties for each cluster and index are listed in
Table~\ref{table_median_uncertainties}.

\begin{table}
\caption{Coefficients of the linear or quadratic fit used to construct the corrected pseudo-indices and marked as
dashed lines in Figures ~\ref{fig_ngc2158}, ~\ref{fig_ngc2420}, ~\ref{fig_ngc2682}, ~\ref{fig_ngc7789}  and ~\ref{fig_be29}, respectively.\label{table_coeff_terms}}
\begin{tabular}{cccc}
\hline\hline
 Index & a & b & c \\
\hline
\multicolumn{4}{c}{NGC~2158}\\
\hline
$S3839_{H}$ & -0.59$\pm$0.19 & 0.004$\pm$0.011 & \\
$S3839_{N}$ & -0.31$\pm$0.22 & 0.001$\pm$0.013 & \\
$S4142$ & -0.22$\pm$0.06 & -0.006$\pm$0.003 & \\
$CH4300$ & -0.60$\pm$0.14 & 0.007$\pm$0.008  & \\
\hline
\multicolumn{4}{c}{NGC~2420}\\
\hline
$S3839_{H}$ & 8.85$\pm$1.74 & -1.22$\pm$0.22 & 0.040$\pm$0.007 \\
$S3839_{N}$ & 9.55$\pm$1.61 & -1.28$\pm$0.20 & 0.042$\pm$0.006 \\
$S4142$ & -2.14$\pm$0.70 & 0.24$\pm$0.09 & -0.008$\pm$0.003 \\
$CH4300$ & 2.23$\pm$0.87 & -0.39$\pm$0.11 & 0.014$\pm$0.003 \\
\hline
\multicolumn{4}{c}{NGC~2682}\\
\hline
$S3839_{H}$ & -13.60$\pm$244.49 & 1.59$\pm$32.65 & -0.05$\pm$1.09 \\
$S3839_{N}$ & -8.56$\pm$13.29 & 0.92$\pm$1.82 & -0.02$\pm$0.06 \\
$S4142$ & 2.81$\pm$9.55 & -0.42$\pm$0.13 & 0.01$\pm$0.04 \\
$CH4300$ & -2.34$\pm$6.27 & 0.22$\pm$0.80 & -0.006$\pm$0.028 \\
\hline
\multicolumn{4}{c}{NGC~7789}\\
\hline
$S3839_{H}$ & 0.11$\pm$3.74 & 0.15$\pm$2.84  & \\
$S3839_{N}$ & 0.45$\pm$3.96 & 0.15$\pm$3.01 & \\
$S4142$ & -1.33$\pm$0.96 & 0.89$\pm$0.73 & \\
$CH4300$ & -0.54$\pm$0.28 & 0.18$\pm$0.22 & \\
\hline
\multicolumn{4}{c}{Berkeley~29}\\
\hline
$S3839_{H}$ & -0.39$\pm$6.90 & -0.00$\pm$0.35 & \\
$S3839_{N}$ & -3.24$\pm$4.16 & 0.15$\pm$0.21 & \\
$S4142$ & -0.00$\pm$1.37 & -0.02$\pm$0.07 & \\
$CH4300$ & -0.98$\pm$1.41 & 0.02$\pm$0.07  & \\
\hline
\end{tabular}
\end{table}

Two or more instrumental configurations were employed to observe three of our clusters: NGC~2420, NGC~2682
and Berkeley~29. Therefore, 102 stars in their regions were observed twice. Although none of them are within those
selected as cluster members, they are very useful for checking the homogeneity of our data and providing an additional estimation of the uncertainties. To do this, each index was measured separately in the two spectra
for each star. On average, the differences between the values obtained for each
index for all the stars are: $\Delta (S3839_N)=0.000\pm 0.005$ ,
$\Delta(S3839_H)=0.000\pm 0.005$ , $\Delta (S4142)=0.003\pm 0.003$ and $\Delta
(CH4300)=0.000\pm 0.003$. These differences are negligible, thus ensuring the homogeneity
of the index determination used in this paper.

\begin{figure}
\centering
\includegraphics[width=\hsize]{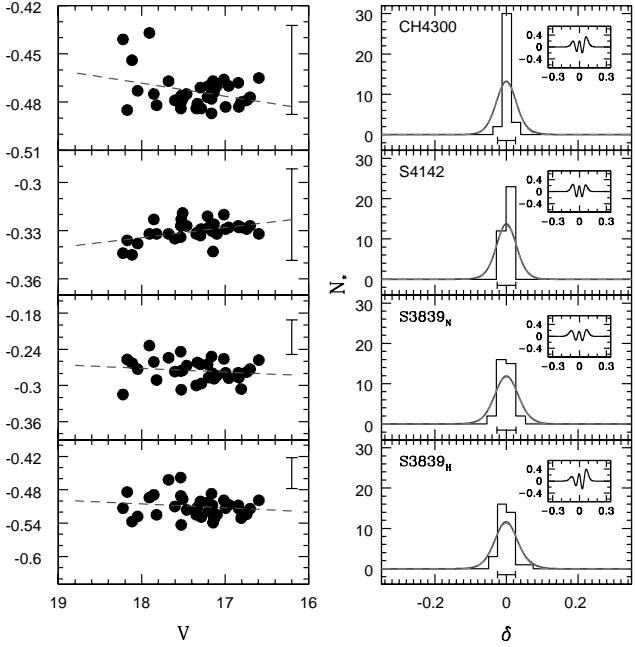}
\caption{Left: run of the strength of the molecular indices of MS stars in NGC~2158 vs.\ their $V$ magnitudes.
A line has been fitted (dashed line) in order to remove the temperature and gravity dependence and in order to
obtain the corrected pseudo-indices. Right: normal (histogram) and
 genralized (solid line) distributions of each
corrected pseudo-index. Dashed lines are the best single Gaussians fitted to each
 generalized distribution.  Note that the dashed lines fall so closely to the solid lines in all the cases that it is difficult to distinguish them. The
residuals between the single Gaussian and the generalized
distribution are shown in the inset panels.\label{fig_ngc2158}}
\end{figure}

The strength of each molecular band depends
not only on chemical abundance but also on the temperature and surface
gravity of the star. Different approaches can be found in the literature to removing
this dependence, such as fitting the lower envelope or the median
ridge line of a given molecular index as a function of colour or magnitude.
In our case, we performed a polynomial least-squares fit on half of the
selected stars (randomly selected) in each cluster. For NGC~2158, NGC~7789 and Berkeley~29 we used a linear
polynomial, while a quadratic one was used in the case of NGC~2420 and NGC~2682, where the selected stars cover
wide magnitude ranges. This procedure was repeated
10$^3$ times using different random subsets each time. The final terms for each index were obtained as the
median of the values obtained in each individual test. With this
procedure we tried to minimize the uncertainties due to the low number of
object studied and the influence of the points on the edges. The final fits adopted in each case are shown as
dashed lines in left panels of Fig.~\ref{fig_ngc2158}, \ref{fig_ngc2420}, \ref{fig_ngc2682}, \ref{fig_ngc7789}
and \ref{fig_be29} and are listed in
Table~\ref{table_coeff_terms}. Finally, we calculated the corrected pseudo-indices, denoted by a $\delta$
preceding the corresponding index, as the difference
between the index and the adopted fit in each case. Histograms were obtained for each corrected index (right
panels of Figures~\ref{fig_ngc2158}, \ref{fig_ngc2420}, \ref{fig_ngc2682}, \ref{fig_ngc7789}
and \ref{fig_be29}).
In each case, generalized histograms (solid lines) were derived by assuming that each
star is represented by a Gaussian probability function centred on its
index value, whose $\sigma$ is equal to the uncertainty in the
determination of the index. The obtained histograms are discussed in depth in the following section.

\section{Cluster-by-cluster discussion}\label{sec4}

\subsection{NGC~2158}

Because of its apparent spheroidal shape, NGC~2158 was considered as a GC in the past. However, it is an
intermediate-age OC \citepads[$\sim$2 Gyr, ][]{2002MNRAS.332..705C} with slightly sub-solar metallicity
\citepads[-0.28, ][]{2011AJ....142...59J} located towards the Galactic anti-centre at a galactocentric distance $R_{gc}\sim 13.5$ kpc. Studies of this cluster have been hampered by the foreground contamination due to its low location in the Galactic plane.

To our knowledge only two studies have obtained chemical abundances for NGC~2158 stars based on high-resolution
 spectroscopic analysis \citepads[see][]{2011A&A...535A..30C}. These two works obtained different iron abundances.
\citetads{2009AJ....137.4753J} obatined [Fe/H]=-0.03$\pm$0.14 whereas \citetads{2011AJ....142...59J} derived
[Fe/H]=-0.28$\pm$0.05. The latter suggests that this difference may be a consequence of the first value
 having been obtained from only one star and by the different temperature scales. Unfortunately, none of these studies provided enough data to investigate the existence of light element anti-correlations among the NGC~2158 stars.

\begin{figure}
\centering
\includegraphics[width=\hsize]{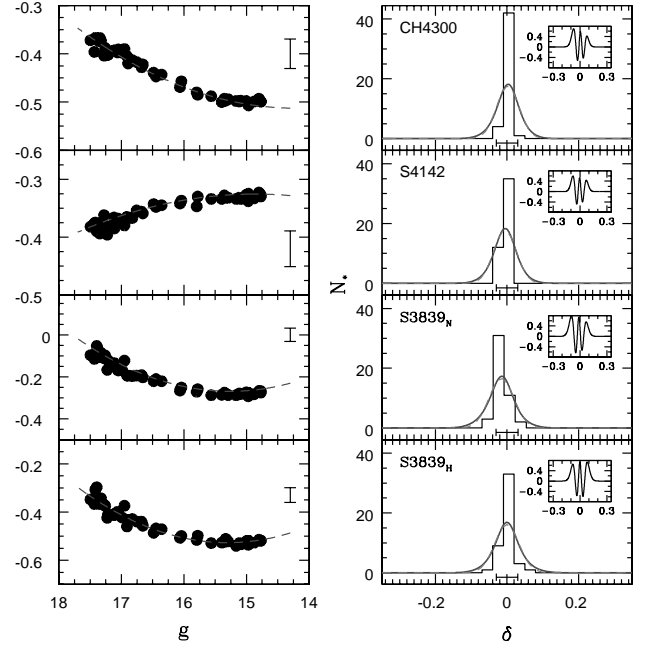}
\caption{As Fig.~\ref{fig_ngc2158} but for NGC~2420.\label{fig_ngc2420}}
\end{figure}

In total, we have measured the CN and CH molecular band strengths for 35 MS stars in this cluster
(Fig.~\ref{fig_ngc2158}). In general, a large scatter is observed in the $V$-index planes (left panels).
This scatter can be explained by a large differential reddening in the NGC~2158 area, as is clearly observed in
its colour--magnitude diagram (Fig.~\ref{fig_dcm}). However, the single Gaussians fitted to the
 generalized histograms of each index (dashed lines in right panels) have sigmas similar to the average uncertainty in each case ($\sim$0.03; see Table~\ref{table_median_uncertainties}). There are no signs of bimodalities or anomalous spreads. The band strengths obtained in our analysis are well reproduced by a single Gaussian distribution  as it is denoted by the residuals between the generalized histograms (solid lines) and the fitted single Gaussians (dashed lines) plotted in inset boxes of the right panels. As in this paper, \citetads{2009PASP..121..577M} analysed the strength of the CN molecular bands for six stars in the upper RGB and RC region. Although their sample is small in comparison with ours, they found no evidence of star-to-star variations among the NGC~2158 stars as in our case.

\subsection{NGC~2420}
NGC~2420 is a $\sim$3 Gyr old intermediate-age OC with a distance modulus of (m-M)$_V$=11.9$\pm$0.3 located
beyond the solar circle at a Galactocentric distance of $\sim$11 Kpc and 0.8 Kpc above the Galactic plane
\citepads[e.g.][]{2010A&A...511A..56P}. It has traditionally been considered among the most metal-poor OCs,
[Fe/H]=-0.57$\pm$0.08 \citepads[e.g.][]{1987AJ.....93..359S}. This has led to NGC~2420 being considered
as a transition system between a solar-metallicity OC and metal-poor GC. However, more recent analysis based on
high-resolution spectra pointed to an iron content slightly above solar. For example,
\citetads{2010A&A...511A..56P} obtained [Fe/H]=-0.05$\pm$0.03 from the analysis of three RC stars,  whereas
\citetads{2011AJ....142...59J} found [Fe/H]=-0.20$\pm$0.06 from the study of nine giants. Although
\citetads{2010A&A...511A..56P} derived abundances for Al, Mg, Na and O, the small number of stars studied
does not allow a proper investigation of the existence of (anti-)correlations between the abundances of these elements.

Although 114 MS stars were selected as members of NGC~2420 according to the criteria described in Section~\ref{sec2}, we have discarded 67 of these because of their low signal-to-noise ratio. We have therefore determined the CN and CH band strengths for 47 MS objects. The run of each of them as a function of $g$ magnitude is shown in the left panels of Figure.~\ref{fig_ngc2420}. Except for the $S4142$ index, it is observed that the strength of the molecular indices decreases towards brighter magnitudes. The same trends are also observed in NGC~2158 (although probably blurred by the differential reddening in the line of sight), NGC~2682 and Berkeley~29. Similar trends were also found in Paper I for NGC~6791. This is explained because for MS stars the temperature increases as the  magnitude decreases, the indices therefore decreasing, since the formation of CN and CH molecules is more efficient at lower temperatures. In contrast, the $S4142$ CN index shows an opposite trend: the strength of this molecular band increases as magnitude decreases. We believe that this trend is artificial since the variation is lower than the average uncertainty in the strength determinations, which is about $\sim$0.03 (error bars in the left panels) for the four indices analysed.

The distributions of the pseudo-indices for NGC~2420 are plotted in the right panels of Figure~\ref{fig_ngc2420}.
Again, there are no signs of bimodalities or dispersions larger than those expected from the uncertainties. The
sigma of the Gaussian fits (dashed lines in the right panels of Fig.~\ref{fig_ngc2420}) is $\sim$0.03, which is
very similar to the average uncertainty of each index (see Table~\ref{table_median_uncertainties}). Note that the solid lines almost cover completely the dashed lines for all the indices studied (right panels of Fig.~\ref{fig_ngc2420}). However, from the residuals shown in the inset panels, it seems that the generalized histograms are not exactly Gaussian. In any case, the symmetrical behaviour of these residuals only implies that the generalized histograms have a higher peak and wider wings that a Gaussian, but there is no evidence of the existence of multiple statistical populations. \citetads{1974ApJ...189..409M} analysed DDO photometry in the $C(41-42)$ bandpass, which includes the CN molecular band at 4142 \AA, for 12 giant stars in NGC~2420. As in our case, they found that all the studied stars have similar $C(41-42)$ magnitudes and therefore, similar CN band strengths.

\begin{figure}
\centering
\includegraphics[width=\hsize]{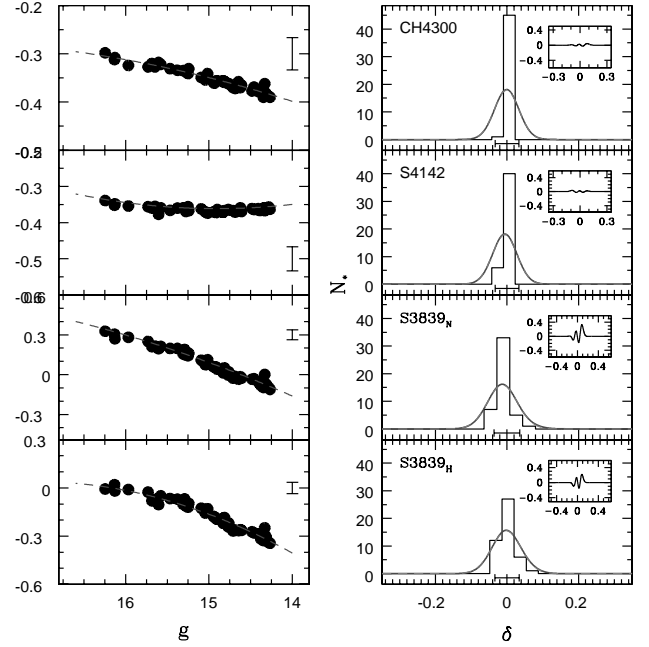}
\caption{As Fig.~\ref{fig_ngc2158} but for NGC~2682.\label{fig_ngc2682}}
\end{figure}

\subsection{NGC~2682 (M~67)}

NGC~2682 is a $\sim$4.5 Gyr old and solar metallicity cluster, [Fe/H]=+0.05$\pm$0.03
\citepads[see][and references therein]{2010A&A...511A..56P}. It is one of the most well
studied OCs due to its low reddening, E(B-V)=0.04$\pm$0.02, and proximity, (m-M)$_V$=9.67$\pm$0.11.
Chemical abundances from high-resolution spectroscopy have been derived by several studies
using different spectral-type stars in NGC~2682 \citepads[see][for a recent compilation]{2011A&A...535A..30C}.
All of them agree on a similar iron content to that of the Sun. Many of these studies have also derived
abundances of light elements such as C, N, Al, Mg, Na and O. There is a wide spread among the
values obtained by different authors for Mg and Na, which may be explained by the difficulties in
measuring these elements, as was pointed out by \citetads{2010A&A...511A..56P}. As in the case of NGC~2420,
these authors derived Al, Mg, Na and O abundances but the small number of stars analysed prevented them from obtaining conclusions about the existence of (anti)-correlations among them.

In this case, we have determined the band strengths of 46 MS stars that show well defined sequences
in the $g$-index planes (left panels of Fig.~\ref{fig_ngc2682}). The median uncertainties in the
determination of the strengths of each index are between 0.03 and 0.04 (Table~\ref{table_median_uncertainties}).
In contrast with the other clusters in our sample, the strengths of the four indices studied increase towards fainter magnitudes, and therefore towards lower temperatures, as expected theoretically
and observed in GCs \citepads[e.g][]{2010A&A...524A..44P} and in NGC~6791 (Paper I). The generalized distributions obtained for each pseudo-index (right panels) are very well reproduced by Gaussians, as the
very small residuals plotted in the inset panels denote. Note that  in right panels of Fig.~\ref{fig_ngc2682} solid and dashed lines overlap. This implies that all the stars analysed in
NGC~2682 have very similar CN and CH band strengths, and therefore C and N abundances, with dispersions
similar to the median uncertainties for each index. The sigmas of the Gaussians fitted to each generalized histogram
are listed in Table~\ref{table_median_uncertainties}. As in the case of NGC~2420, our results are in good agreement with the conclusions obtained from DDO photometry in 36 red giant stars studied by \citetads{1984AJ.....89..487J}.

\begin{figure}
\centering
\includegraphics[width=\hsize]{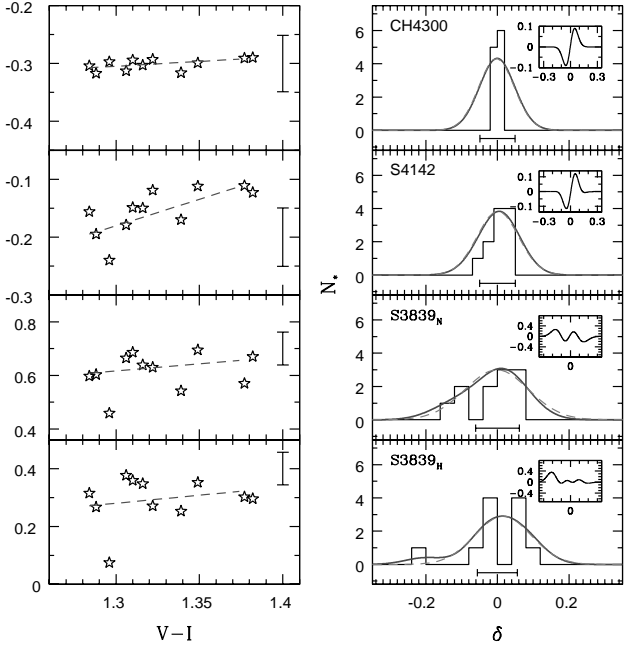}
\caption{As Fig.~\ref{fig_ngc2158} but for RC stars in NGC~7789. In this case the run of each index has been plotted as a function of $V-I$ colours.\label{fig_ngc7789}}
\end{figure}

\subsection{NGC~7789}

NGC~7789 is a $\sim$1.5 Gyr old OC located at a Galactocentric distance of $R_{gc}\sim$9.4 kpc and
about 0.2 kpc below the Galactic disc plane, which explains the relatively high reddening in its line
of sight, E(B-V)=0.27$\pm$0.04 \citepads[see][and references therein]{2010A&A...511A..56P}. Although
studies based on photometry and low resolution spectroscopy assign a sub-solar metallicity to this cluster,
[Fe/H]$\sim$-0.24$\pm$0.09 \citepads[e.g.][]{2002AJ....124.2693F}, high-resolution spectroscopic studies have found an iron content similar to that of the Sun: [Fe/H]=+0.04$\pm$0.07
\citepads{2010A&A...511A..56P}. Although \citetads{2010A&A...511A..56P} derived  Al, Mg, Na and O abundances in three RC stars, the small number of objects studied does not allow the existence of usual (anti)-correlations observed in GCs to be investigated.

\begin{figure}
\centering
\includegraphics[width=\hsize]{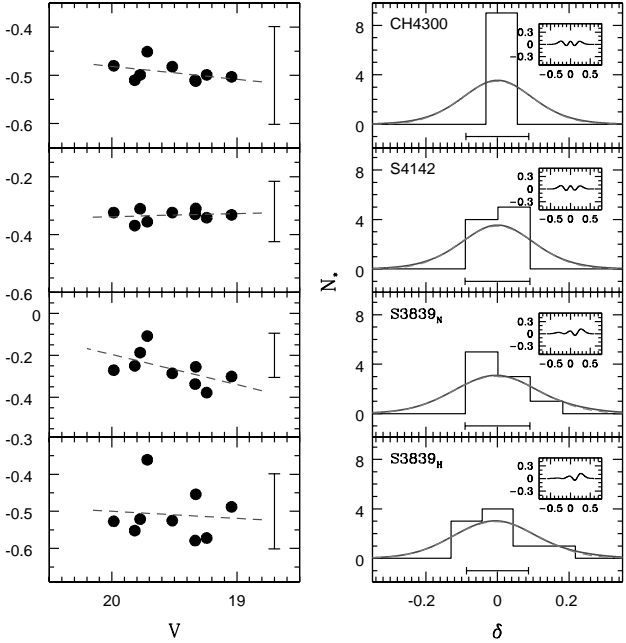}
\caption{As Fig.~\ref{fig_ngc2158} but for Berkeley~29.\label{fig_be29}}
\end{figure}

In contrast with the other clusters in our sample, we have restricted our analysis of NGC~7789 to 11 RC stars because no MS objects were observed in this cluster. Since RC objects are located almost at the
same luminosity, it is not useful to use the magnitude to remove the temperature and gravity dependences.
For this reason in the left panels of Figure~\ref{fig_ngc7789} we have plotted the run of the index strengths
as a function of $V-I$ colour. Although the uncertainties in the four indices are relatively large (see
Table~\ref{table_median_uncertainties}), the expected increase of the band strengths towards redder colours,
and therefore towards lower temperatures, is observed in all of them. The $CH4300$ index shows a unimodal
distribution within the uncertainties, as has been observed in the other clusters studied here. However, the $S4142$ index shows a generalized histogram that, although it is relatively well fitted by a single Gaussian, has a sigma that is slightly larger than the average uncertainty. Hints of bimodalities are observed in the histograms of both $S3839_H$ and $S3839_N$ - particularly in the latter, which was defined to sample giant stars such as on the RC. However, these bimodalities are smoothed when the generalized distributions are obtained taking into account the uncertainties. However, none of them is well reproduced by a single Gaussian as the residuals plotted in inset boxes of the right panels indicate. This broadening is not explained by the uncertainties since the sigmas obtained from the single Gaussian fits are larger than the average error bars (see Table~\ref{table_median_uncertainties}). We have also investigated the possibility that the observed scatter is independent of the $V$ magnitude, i. e. if the stars analysed are in different evolutionary stages. The star with the lower $S3839_H$ index is the faintest one with $V$=13.4. There are no other stars in our sample with similar magnitudes. Therefore, its lower CN band strengths could be explained by the fact that this star would be in a different evolutionary stage that the others. In contrast, for the other stars analysed, we find that the objects with lower CN band strengths have almost the same magnitudes than others with higher values. We have also investigated if our result is influenced by the existence of binaries in our sample. According to the WEBDA database only two stars in our sample are spectroscopic binaries These stars, 8383, 9831, fall in the most populated part of the V-I vs indices diagrams at around $V-I\sim$1.32. The unexpected broadenings observed in the generalized histograms of $S3839_H$ and $S3839_N$ indices do not disappear if these stars are removed from the sample. Therefore, we conclude that our result is not explained by the existence of stars in different evolutionary stages or binaries in our sample. The spectra of two stars with similar $V-I$ colours but different $S3839_H$ and $S3839_N$ strengths are shown in Figure~\ref{fig_spectra}. A similar behaviour was observed in NGC~6791 RC stars in Paper I. However, both the small number of stars sampled and the relatively low signal-to-noise ratio of the spectra used in our analysis, which implies large uncertainties, prevent us from drawing further conclusions. In any case, although we cannot confirm the existence of star-to-star variation of CN band strengths, neither can we discard them. DDO photometry were obtained by \citetads{1977AJ.....82...35J} for 22 giants in NGC~7789. From the analysis of the magnitudes obtained in the $C(41-42)$ bandpass they did not find evidences of an intrinsic range of CN strengths in contradiction with the result obtained here. However, the lower sensitivity of CN molecular band at 4142 \AA~to the star-to-star variations than the one located at 3839 \AA~may explain this discrepancy.

\subsection{Berkeley~29}

Berkeley~29 is a very interesting cluster because of its location towards the Galactic anti-centre in the
outskirts of the Galactic disc at a Galactocentric distance of $R_{gc}\sim$23 Kpc and height of $\sim$2 Kpc
above the disc. Its intermediate-age, $\sim$3.5 Gyr, and low metallicity, [Fe/H]$\sim$-0.4
\citepads[e.g.][]{2004MNRAS.354..225T,2005A&A...429..881B} match quite well within the framework of
Galactic evolution in spite of the suggestion by \citetads{2004ApJ...602L..21F} that Berkeley~29 is associated
with the Canis Major overdensity. In any case, Berkeley~29 is a very interesting object; we have therefore included it in our sample in spite of the low number of stars selected as members in comparison with the other systems analysed, except for NGC~7789.

Of the 20 objects selected as Berkeley~29 members from the criteria described in Section~\ref{sec2} we have
determined CN and CH band strengths for only nine MS stars. Their run as a function of $V$ magnitudes are shown
in left panels of Figure~\ref{fig_be29}. The trends are similar to those observed in other clusters in our
sample, although the scatter is larger because of the larger uncertainties motivated by the lower signal-to-noise
ratios of the spectra acquired for Berkeley~29 stars. Moreover, the generalized histograms for each pseudo-index are well reproduced by single Gaussians with sigmas similar to the median uncertainties in each case (see
Table~\ref{table_median_uncertainties}). This implies that the generalized histograms are mainly modulated
by the large uncertainties in the determinations of the four molecular indices analysed, which prevents us from obtaining any conclusion about the existence of star-to-star variations. We therefore conclude that, within
the uncertainties, we find no evidence of the existence of significant differences in the strengths of
the CN and CH molecular bands, and therefore in the abundances, of the analysed stars.

\section{Discussion}\label{sec5}

\begin{table}
\caption{List of previously studied open clusters\label{tbl-X}}
\centering
\begin{tabular}{lcccc}
\hline\hline
Cluster & Stars & Type & Ref. \\
\hline
NGC~188 & 13 & MS & \citetads{1985AJ.....90.2526N} \\
NGC~188 & 7 & RC/RGB & \citetads{2009PASP..121..577M}\\
NGC~2158 & 6 & RC/RGB & \citetads{2009PASP..121..577M} \\
NGC~2158 & 10 & RC & \citetads{1984AJ.....89..263S} \\
NGC~6791\tablefootmark{1} & 31 & RC/RGB & \citetads{1995AJ....110..693H}\\
NGC~6791 & 97 & MS/RGB/RC & Paper I \\
NGC~7789 & 10 & RC/RGB & \citetads{2009PASP..121..577M} \\
\hline
\end{tabular}
\tablefoottext{1}{Although \citetads{1995AJ....110..693H} found evidence of variations in the CN band
strengths among the stars in their sample, they did not associate them with the existence of star-to-star abundance variations.}
\end{table}

The existence of multiple populations has been reported in massive clusters with a wide range of ages,
from old  GCs to intermediate-age and young populous systems in the Magellanic Clouds. They are
clearly linked with the star-to-star variations of light element chemical abundances reported in almost
all GCs properly studied not only in the Galaxy but also in the LMC and Fornax dwarf spheroidal galaxy. However, the same chemical inhomogeneities have not been observed in intermediate-age populous clusters of the LMC \citepads{2008AJ....136..375M,2011MNRAS.415..643W}. Whatever processes may be responsible for the formation of multiple populations in stellar clusters, it appears that it is mainly related to their total mass since only massive enough systems are able to retain the material ejected by the first generation, from which the subsequent populations are formed. Thus, there should be a mass limit below which
 clusters are unable to retain the ejected material and therefore to form other stellar generations.
 According to dynamical models, this mass threshold is between $\sim$ 10$^5$ and 10$^6$ M$_{\odot}$
 \citepads[e.g.][]{2010ApJ...718L.112V,2011MNRAS.412.2241B}.

Constraining this limit from an observational point of view is challenging since clusters lose a significant
fraction of their initial masses during their lifetimes not only as a function of stellar evolution
(e.g. material removed by supernova explosions or strong winds) but also by tidal stripping. However,
 investigating less massive clusters helps to restrict the minimum mass required. In this paper we have
studied the behaviour of the CN and CH molecular band strengths in five OCs. Four of them (NGC~2158,
NGC~2420, NGC~2682 and Berkeley~29) have unimodal CN and CH band distributions within the uncertainties,
suggesting that they harbour a simple and chemically homogeneous population. In the case of the most massive
of these clusters, Berkeley~29, the uncertainties are large enough to erase any sign of either bimodalities or
broadening in the obtained distributions. In contrast, for NGC~7789, with a mass similar to that of NGC~6791,
we detected an anomalous spread in the distribution of the CN band strengths in spite of the large
uncertainties in the index determinations for stars in this cluster.

\begin{figure}
\centering
\includegraphics[width=\hsize]{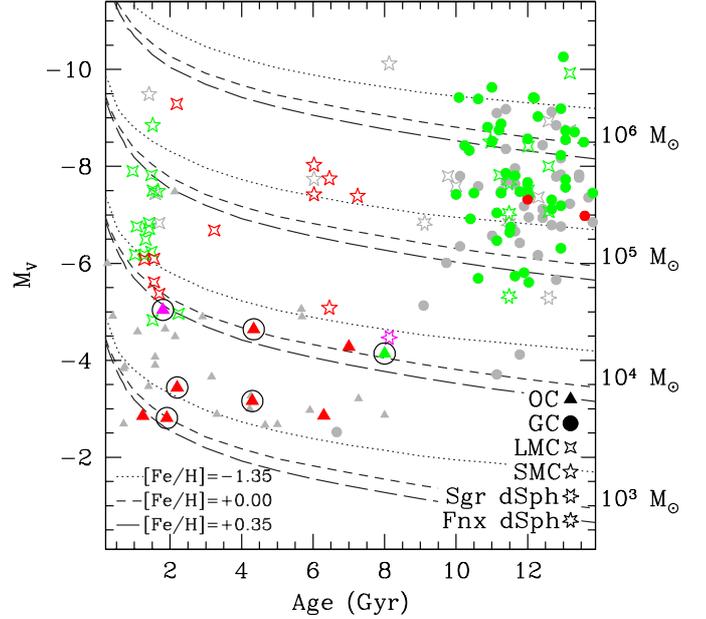}
\caption{Age versus M$_V$ (as a proxy for mass) for Galactic OCs (triangles) and GCs (filled circles).
Clusters in other galaxies have been marked with stars with different numbers of points: LMC (four), Small Magellanic
Cloud (SMC, five), Sagittarius (six) and Fornax (seven). Clusters for which either multiple sequences in the
colour--magnitude diagrams or chemical inhomogeneities, or both, have been detected are in green whereas
clusters without either are in red. In magenta are clusters for which there are doubts about the existence of
chemical inhomogeneities. Clusters not properly studied are in grey. Those clusters studied here are encircled. Lines of constant mass for different
metallicities have been superimposed (see text for details).
\label{fig_discuss}}
\end{figure}

To put our results in a general context, we used the age--M$_V$ plane originally proposed by
\citetads{2010A&A...516A..55C}, where M$_V$ is used as an indicator of the present-day total mass of the
clusters. In Figure~\ref{fig_discuss} we have plotted several OCs with filled triangles. Those studied
here and in Paper I are  marked with open circles. NGC~6791, the only OC where the existence of
star-to-star variations of light element abundances have been confirmed,  has been coloured in green.
 NGC~7789 has been coloured in magenta since we have found evidence of the existence of chemical
  inhomogeneities in its stars. Systems with no signs of multiple populations, including those studied here
 and others previously investigated in the literature (Table~\ref{tbl-X}), are coloured in red. Other
 interesting OCs not studied yet are in grey. Star-to-star variations of light element abundances have been investigated in several OCs \citepads[e.g.][]{2009A&A...500L..25D,2009A&A...502..267S,2010A&A...511A..56P,2011A&A...535A..30C},
 however, the number of stars analysed in each system is very small and these have not been considered in our
 analysis. Following the same colour code we have also plotted Galactic GCs (filled circles). Clusters in external galaxies have been plotted with open stars with different numbers of points: LMC (four), Small Magellanic
 Cloud (SMC, five), Sagittarius (six), and Fornax (seven). These clusters and the references used are
 listed in Tables~6, 7 and 8. Lines of constant mass for different metallicities have been superimposed for reference. They have been computed from the simple stellar population models derived by
 \citetads{1998MNRAS.300..872M,2005MNRAS.362..799M} assuming the \citetads{2001MNRAS.322..231K}
 initial mass function\footnote{Available at {\tt http://www.icg.port.ac.uk/~maraston/Claudia's\_Stellar\_Population\_Model.html}}.

No evidence for the existence of multiple populations has been reported for any of the five Galactic
OCs studied with M$_V<$-4. In particular, we have analysed a statistically significant number of stars
in three systems in this magnitude range (NGC~2158, NGC~2420 and NGC~2682) and we do not find any hint
of chemical inhomogeneities.  We conclude that clusters with M$_V<$-4 or masses lower than
$\sim$10$^4$ M$_{\odot}$ are not massive enough to form a second stellar generation.

Independently of their parent galaxy, almost all clusters older than $\sim$9 Gyr are massive enough to
host multiple stellar populations. Moreover, almost all these systems show star-to-star variations of light element abundances. The exceptions are the two Galactic GCs NGC~5466 and IC~4479, where no anomalous colour spreads in the ultraviolet have been detected by \citetads{2011A&A...525A.114L} or \citetads{2011MNRAS.415..643W}. However, a similar result was found by \citetads{2011A&A...525A.114L} in
NGC~2419, but later an unnaturally wide RGB was found by \citetads{2013MNRAS.431.1995B} using higher
quality photometry. To our knowledge the existence of star-to-star chemical abundance variations has not been studied until now in these two clusters.

SMC clusters with ages between 6 and 8 Gyr do not show unnatural features in their colour--magnitude
diagrams in spite of their being massive enough, M$_V\sim$-8, to host multiple stellar populations.
However, these features have been observed in clusters with ages younger than $\sim$2.5 Gyr, not only
in the SMC \citepads{2008AJ....136.1703G} but also in the LMC \citepads[e.g.][]{2009A&A...497..755M,2013MNRAS.430.2358P}.
There are some exceptions, such as the massive cluster SL556 (M$_V\sim$-9.5) in the LMC
\citepads[e.g.][]{2003AJ....125..770B}. In contrast, star-to-star variations of light element
abundances similar to those observed in GC have not been observed in intermediate-age LMC clusters
\citepads{2008AJ....136..375M,2011MNRAS.413..837M}. However, almost all the clusters analysed
 (NGC~1651, NGC~1866 and NGC~2173) also show no unnatural spreads in the MS turn-off region. The only exception is NGC~1783, where although neither an anomalous MS turn-off \citepads{2009A&A...497..755M} nor chemical inhomogeneities \citepads{2008AJ....136..375M} have been initially reported, a more accurate analysis of
the MS turn-off region has revealed an unnatural spread \citepads{2011ApJ...733L...1P}. In any case, more
determinations of chemical abundances of stars in these clusters, and in particular in those with unnatural or
multiple MS turn-offs, are needed to confirm that clusters younger than $\sim$2.5 Gyr are able to form
multiple populations but without the chemical inhomogeneities observed in old GC stars.

According to Figure~\ref{fig_discuss}, the mass limit for forming multiple populations is between M$_V\sim$-4 and -5. Unfortunately, the number of clusters studied in this range is small. Unnatural MS turn-offs have been
observed in two LMC young clusters in this magnitude range, SL~529 and SL~862 by \citetads{2013MNRAS.430.2358P} and \citetads{2009A&A...497..755M}, respectively. However, they have not been reported in slightly more massive clusters at M$_V\sim$-5.5. The only GC in this range is Palomar~12, which is associated with the Sagittarius dwarf spheroidal galaxy. As was explained in Section~\ref{sec1}, star-to-star variations of light element abundances derived from high resolution spectroscopy have not been reported in this
cluster \citepads{2004AJ....127.1545C}. However, only four stars have been analysed. In the same way,
\citetads{2008A&A...486..437K} find no significant variations of the CN band stengths in their
analysis of low-resolution spectra of 14 RGB stars in Palomar~12. In contrast, a clear bimodal anti-correlation
of the CH and CN band strengths has been observed by \citetads{2010A&A...524A..44P} by analysing 23
stars in the MS turn-off region. Since there is no agreement about the existence of chemical inhomogenities
in Palomar~12 we have coloured this cluster in magenta in Figure~\ref{fig_discuss}. The other four systems
in this magnitude range are OCs. Both chemical inhomogeneities in light element abundances derived from
high-resolution spectroscopy \citepads{2012ApJ...756L..40G} and variations of the CN band strength measured
in low-resolution spectra (Paper I) have been reported in the metal-rich OC NGC~6791. Moreover, this cluster
shows unnaturally wide RGB and MS sequences \citepads{2006ApJ...643.1151C}. However, star-to-star variations
have not been observed in Berkeley~39 \citepads{2012A&A...548A.122B}. Although both systems have a similar
present-day mass, it seems that NGC~6791 had a higher initial mass than Berkeley~29. NGC~6791 would have lost
a significant fraction of its initial mass during repeated crossings of the denser parts of the disc. The
remaining OCs are those studied here: NGC~7789 and Berkeley~29. Only nine stars have been analysed in the case of
Berkeley~29. The uncertainties of the index determinations are large, which would erase any sign of bimodality
or anomalous spread because of the low signal-to-noise ratio of the spectra used. In contrast, in the case of
NGC~7789 we found an anomalous spread, not explained by the uncertainties, in the distribution of CN band strength at 3839 \AA~(Fig.~\ref{fig_ngc7789}). Although further studies are needed to confirm this result, it has a similar age and mass as the two LMC clusters where unnatural MS turn-offs have been reported.

Although much work remains to be done, according to Figure~\ref{fig_discuss} the minimum present-day mass required for a cluster to be able to retain the material ejected by the first stellar generation and form subsequent
populations in $\sim$10$^4$ M$_{\odot}$. There are some clusters with a present-day mass similar to this that do
not host multiple stellar populations. This may be because they have lost a low amount of mass during their lifetimes in comparison with other systems with similar present-day mass that were initially more massive. This value is lower than the mass threshold of 10$^4$-10$^5$ M$_{\odot}$ proposed by \citetads{2009ApJ...695L.134M,2010A&A...516A..55C}.

\section{Summary}

We have studied the strengths of the CN and CH bands at 3839, 4142 and 4300 \AA~in MS stars of NGC~2158, NGC~2420, NGC~2682 (M~67) and Berkeley~29, and RC objects of NGC~7789. For this analysis, we used low-resolution spectra (R$\sim$2000) obtained in the framework of the SEGUE project within the SDSS. Our main results are as follows:

\begin{itemize}
\item The molecular indices describe unimodal distributions for stars in NGC~2158, NGC~2420, NGC~2682 and
Berkeley~29. All of them are well reproduced by single Gaussians with sigmas similar to the average uncertainties.
Therefore, stars in these clusters have homogeneous abundances of C and N.
\item Hints of bimodalities have been found in the distributions of the strength of the CN molecular band at
3839 \AA~for stars in NGC~7789. Although these features are smoothed when the uncertainties are taken into
account, the generalized histograms still show a spread wider than that expected from the error bars. Further
analysis is needed to confirm the existence of star-to-star variations in the CN strengths of stars in this
cluster because of the small number of objects studied here.
\item We have discussed our results in the framework of the formation of multiple stellar populations in
stellar clusters of different ages and environments. We find that the present-day mass limit above which a
cluster may be able to host multiple stellar populations is $\sim$10$^4$ M$_{\odot}$.
\end{itemize}

\begin{acknowledgements}
CEMV acknowledges the funds by the Instituto de Astrof\'{\i}sica de Canarias under the Summer
Grant Programme. R.C. acknowledges funds provided by the Spanish Ministry of Science and
Innovation under the Juan de la Cierva fellowship and under the Plan Nacional de
Investigaci\'on Cient\'\i fica, Desarrollo, e Investigaci\'on Tecnol\'{\i}gica, AYA2010-16717.
This research has made use of the WEBDA database, operated at the Institute for Astronomy of the
University of Vienna, and the SIMBAD database, operated at CDS, Strasbourg, France.

Funding for SDSS-III has been provided by the Alfred P. Sloan Foundation, the
Participating Institutions, the National Science Foundation, and the U.S.
Department of Energy Office of Science. The SDSS-III web site is
http://www.sdss3.org/. SDSS-III is managed by the Astrophysical Research
Consortium for the Participating Institutions of the SDSS-III Collaboration
including the University of Arizona, the Brazilian Participation Group,
Brookhaven National Laboratory, University of Cambridge, University of Florida,
the French Participation Group, the German Participation Group, the Instituto de
Astrof{\'i}sica de Canarias, the Michigan State/Notre Dame/JINA Participation Group,
Johns Hopkins University, Lawrence Berkeley National Laboratory, Max Planck
Institute for Astrophysics, New Mexico State University, New York University,
Ohio State University, Pennsylvania State University, University of Portsmouth,
Princeton University, the Spanish Participation Group, University of Tokyo,
University of Utah, Vanderbilt University, University of Virginia, University of
Washington, and Yale University.\end{acknowledgements}

\end{document}